\documentclass[useAMS,usenatbib]{mn2e}
\usepackage{graphicx}
\usepackage{url}

\newcommand{\etal}{\textit{et al.}}
\newcommand{\lcdm}{$\Lambda$CDM}

\title[Testing for Anisotropic Dark Energy]{Does the Universe Accelerate Equally in all Directions?}
\author[R. Cooke and D. Lynden-Bell]{Ryan Cooke$^{1}$\thanks{E-mail:
rcooke@ast.cam.ac.uk; dlb@ast.cam.ac.uk} and Donald Lynden-Bell$^{1}$\\
$^{1}$Institute of Astronomy, Madingley Road, Cambridge, CB3 0HA}
\begin{document}

\date{Accepted 2009 September 20.  Received 2009 September 11; in original form 2009 August 1}

\pagerange{\pageref{firstpage}--\pageref{lastpage}} \pubyear{2009}

\maketitle

\label{firstpage}

\begin{abstract}

We employ the Union compilation of Type Ia supernovae (SNe) with a maximum likelihood analysis to search for a dark energy dipole. To approach this problem, we present a simple, computationally efficient, and largely model independent method. We opted to weight each SN by its observed error estimate, so poorly measured SNe that deviate substantially from the Hubble law do not produce spurious results. We find, with very low significance, a dipole in the cosmic acceleration directed roughly towards the cosmic microwave background (CMB) dipole, but this is almost certainly coincidental.

\end{abstract}


\section{Introduction}
The foundations for the Friedmann-Lema\^itre-Robertson-Walker (FLRW) cosmological model rest on the assumptions of isotropy and homogeneity, both of which seem plausible on average. However, given the discovery of cosmic acceleration \citep{rie98,per99}, the FLRW model must attribute this observation to a cosmological constant (dark energy) in Einstein's equations, which still lacks an acceptable physical interpretation (see \citealt*{fri08} for a review).

To alleviate these concerns, numerous authors have investigated possible inhomogeneous, and void models to account for the observed acceleration (see \citealt{enq08} for a review, but also \citealt*{aln06} and references therein). The typical approach is to employ the exact Lema\^itre-Tolman-Bondi solutions for a radially inhomogeneous, spherically symmetric Universe, and require that we must be almost centrally located in a large scale void. However, a need to violate the Copernican principle has not yet been evidenced. Moreover, many of these models seem unlikely (see \citealt{ish06}).

A more recent development by \citet{wil07a}, suggests that cosmic acceleration can be explained in general relativity, by considering the differences in the quasi-local gravitational energy between observers in bound systems (clusters of galaxies in bubble walls and filaments) and those in a freely expanding space (volume-averaged observers), resulting in a cumulative time dilation effect due to the differences in their clock rates (see also \citealt{wil07b}). Using the Gold SNe sample \citep{rie07}, \citet*{lei08} demonstrated that this model provides a possible alternative to dark energy, although this model fails to provide a consistent fit to the Union \citep{kow08} and Constitution \citep{hic09} SNe samples \citep*{kwa09}, contrary to the current concordant cosmology (\lcdm), whereby all three SNe samples are fit by a consistent set of cosmological parameters.

Other authors have investigated the possibility that dark energy is anisotropic (see \citealt*{coo08} and references therein). \citet{coo08} present a viable method to model dark energy inhomogeneities as a power spectrum of luminosity fluctuations. However, with such optimistic objectives, these studies will require a dedicated all-sky survey to search for SNe. One can, however, apply a simple method to measure the anisotropy of dark energy, or alternatively, search for a systematic directional dependence in the SNe data \citep*{kol01,gup08}. This involves dividing the data into two hemispheres, and searching for the strongest directional dependence, which these authors find to have a significance of about $1\sigma$ for previous compilations of SNe.

In this article we employ the current SNe dataset and apply a new technique to test for the presence of a dark energy dipole. In \S2 we detail the SN sample used in our analysis, and then formulate our dipolar model in \S3. In \S4 we present the results of our dipolar modelling, and summarise our conclusions in \S5. Throughout we adopt the cosmological parameters derived by \citet{dun09}; $H_{0}=71.9$\,km\,s$^{-1}$\,Mpc$^{-1}$, $\Omega_{m}=0.258$, and $\Omega_{\Lambda}=0.742$.

\section{The Type Ia Supernova Sample}

\begin{figure}
\includegraphics[width=8cm]{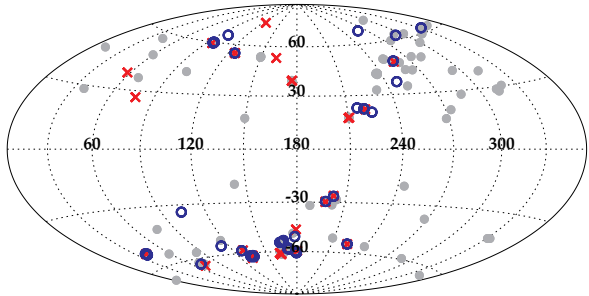}
\caption{
A projection of the spatial distribution of the Union SNe sample in Galactic coordinates.
Note the relative uniformity of the points, except around the Galactic plane. The symbols correspond
to those in Fig.~\ref{fig:hubdev}, and are explained in \S\ref{sec:hubdev}.
}
\label{fig:hubdist}
\end{figure}

For the present analysis, we employ the Union sample of SNe, compiled by \citet{kow08}\footnote{Available from: \url{http://supernova.lbl.gov/Union/}}. The Union compilation contains $307$ identically processed SNe, drawn from a heterogeneous sample of $414$. For the present work, it is noteworthy that \citet{kow08} corrected their data for the CMB dipole, and rejected SNe with CMB-centric redshifts\footnote{\citet{kow08} find that their results do not depend significantly on the adopted redshift cutoff.} $z<0.015$. The Galactic coordinates for all SNe in the Union sample were obtained from the NASA/IPAC Extragalactic Database (NED)\footnote{\url{http://nedwww.ipac.caltech.edu/}}. Indeed, as illustrated in Fig.~\ref{fig:hubdist}, the Union SNe compilation are drawn from a spatially uniform sample, except of course the trivial lack of SNe observed in the Galactic plane.

We acknowledge the presence of a more recent sample of low redshift SNe \citep{hic09}, and decide against including these additional data for the following two reasons. First, the signature of an anisotropic acceleration of the Universe is only detectable at higher redshifts, where the effect of a cosmological constant is more exposed, and the motion of the local group is less of a contaminant. Therefore a low redshift sample should not affect the final result. Moreover, for this reason, we opted to include only the Union SNe with $z>0.2$, leaving us with a total sample of 250. Second, the more recent data are processed using different routines, which could introduce a bias if we were to blindly combine both samples. With the goal of the present work borne in mind, we restrict our analysis to the Union compilation.

\section{Analysis}

As outlined by \citet{kol01}, a number of sources could produce deviations from isotropy in the SN data. From the results of their study, these authors raised concerns about the relatively small number of SNe, spread inhomogeneously across the sky, with differing and uncertain errors. We therefore seek a method that is not biased by the clustering of the SNe data, whilst taking into consideration the individual uncertainties of each SN. Bearing in mind these characteristics, we employ a maximum likelihood strategy (unbiased to clustering), and devise a method that appropriately weights the contribution of each SN based on its corresponding error measurement.

\subsection{The Hubble Deviation}\label{sec:hubdev}
\begin{figure}
\includegraphics[width=8cm]{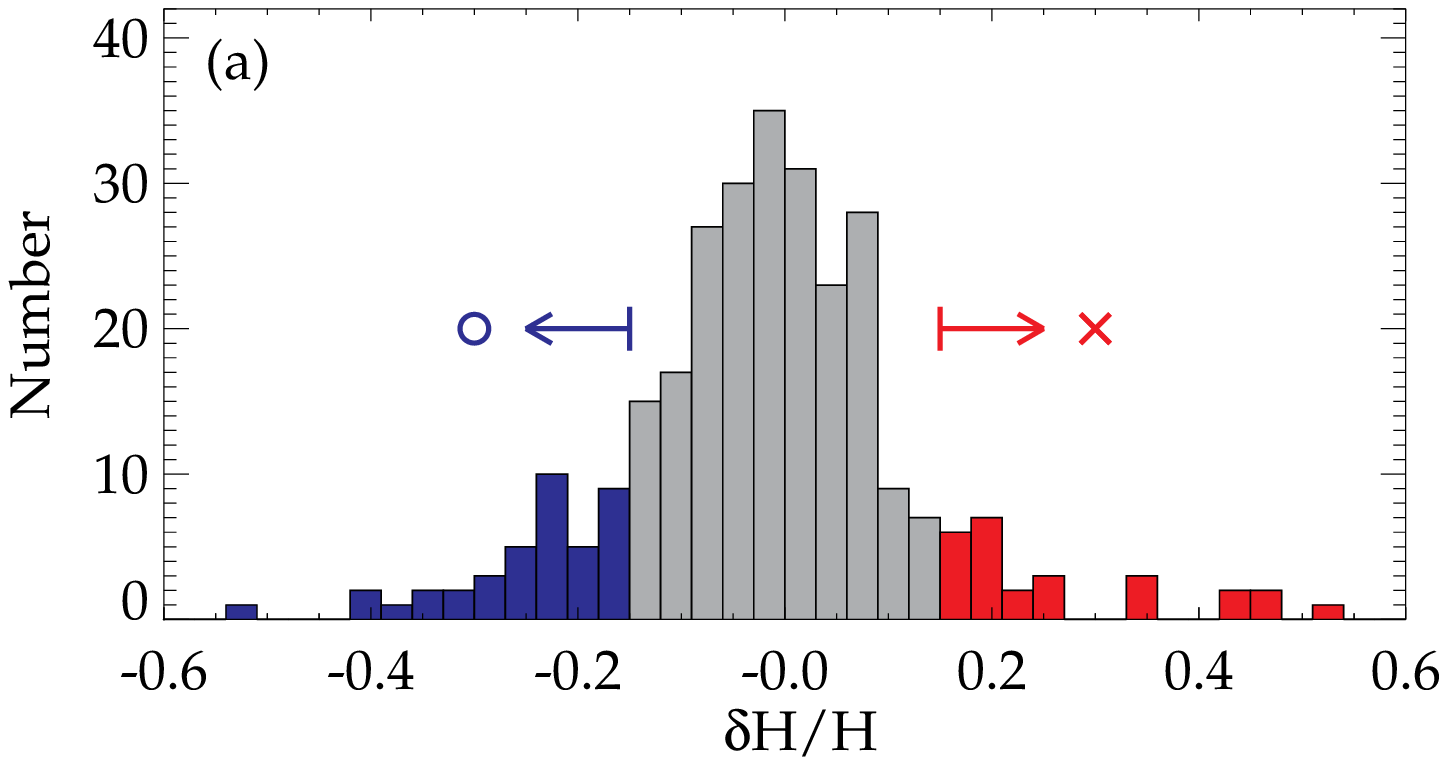}
\newline\newline
\includegraphics[width=8cm]{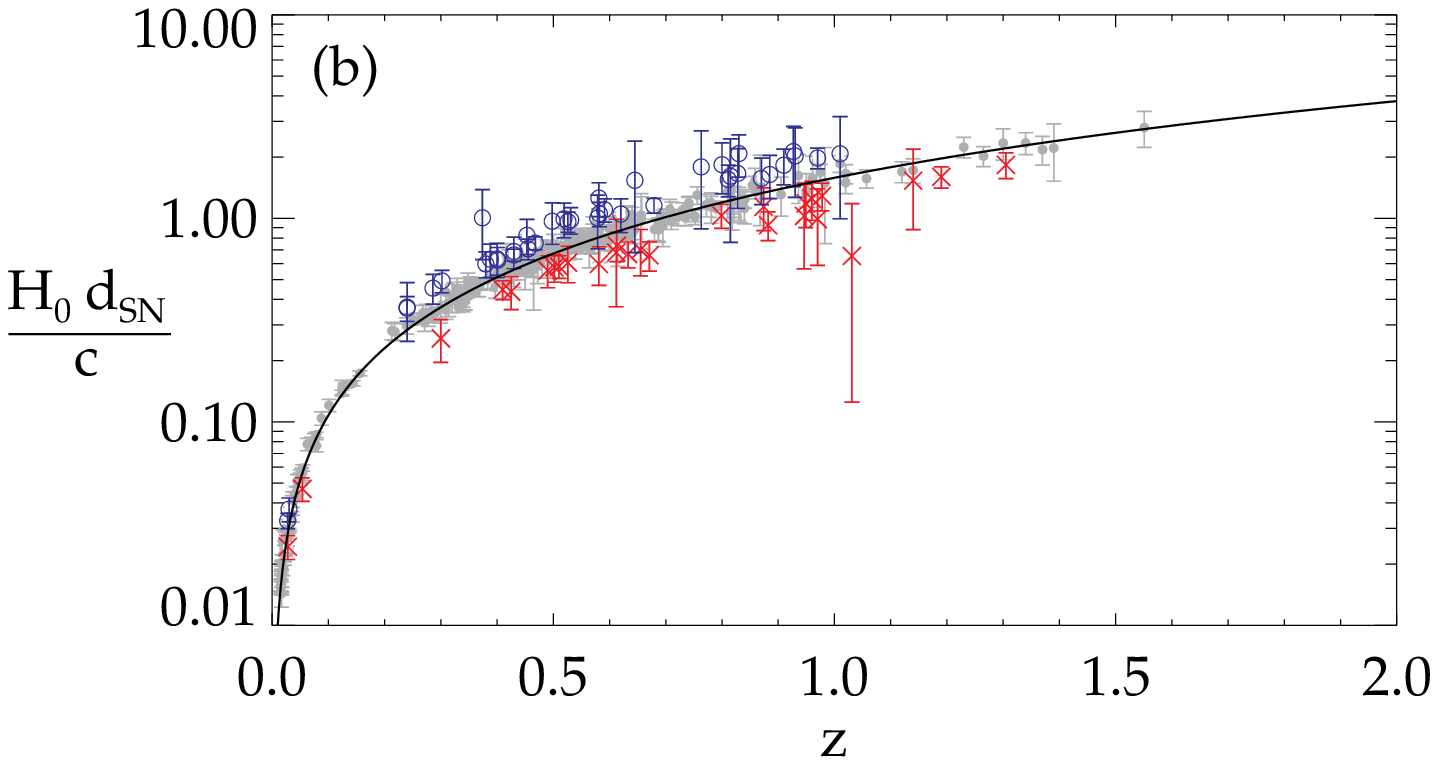}
\caption{
(a) Histogram of deviations from the Hubble law.
(b) A Hubble plot of the entire Union SNe sample (grey symbols), and those with the largest deviations (crosses/circles coloured red/blue represent positive/negative deviation). The solid line corresponds to the Hubble law for the concordance cosmology.}
\label{fig:hubdev}
\end{figure}

Before proceeding with a quantitative analysis of the Union sample, it is instructive to
inspect whether any obvious preferred deviations from the Hubble law exist in the employed dataset.
Following \citet*{jha07}, the deviation from the Hubble law is given by

\begin{equation}
\frac{\delta H}{H}=\frac{H_{0}\,d_{L} - H_{0}\,d_{SN}}{H_{0}\,d_{SN}}\label{eqn:hubdev}
\end{equation}
where $d_{L}$ is the luminosity distance in a $k=0$ Universe,
\begin{equation}
d_{L}=\frac{(1+z)c}{H_{0}}\int_{0}^{z} [\Omega_{\Lambda}+\Omega_{M}(1+z')^{3}]^{-1/2}\,\mathrm{d}z'
\end{equation}
and $d_{SN}$ can be calculated from the observed distance modulus, $\mu=5\log d_{SN} + 25$. Fig.~\ref{fig:hubdev}\,(a) illustrates that the data are only very slightly skewed; more SNe are above the Hubble law (i.e. with negative Hubble deviation) than below (positive Hubble deviation). Moreover, perhaps unsurprisingly, there are also a greater number of SNe that lie above the Hubble law when considering the subsample of SNe that exhibit the largest Hubble deviations ($\,\mid\!\!\delta H/H\!\!\mid\,\,>\,0.15$), represented in Fig.~\ref{fig:hubdist} and \ref{fig:hubdev} by the crosses and open circles. The entire SNe sample are presented on a Hubble plot in Fig.~\ref{fig:hubdev}\,(b). The fact that all of the largest deviations are exhibited by SNe at $z>0.2$, validates our decision not to include SNe with $z<0.2$.

The spatial distribution of the entire Union sample of SNe is presented in Fig.~\ref{fig:hubdist}, where the symbols correspond to those in Fig.~\ref{fig:hubdev}. Indeed, from this simple demonstration, an \emph{obvious} dipole is not apparent.

\subsection{Maximum Likelihood Strategy}\label{sec:max_like_strategy}
We consider a $k=0$ Universe, and compare two models with and without an acceleration term. In essence, the result should not be affected by the model chosen to include the acceleration term, provided that the model represents the centroid of the data at a given redshift (i.e. selecting the \lcdm\ model is sufficient). We define $\delta(\Lambda)=d(\Lambda)-d(\Lambda=0)$ to be the luminosity distance deviation of an isotropically accelerating Universe, $d(\Lambda)$, from a non-accelerating Universe, $d(\Lambda=0)$. The luminosity distance we would measure in a Universe with an anisotropic acceleration $d(\Lambda_{A})$, can then be simply described by

\begin{equation}
d(\Lambda_{A}) = d(\Lambda) + \delta(\Lambda)\,X \label{eqn:ML01}
\end{equation}
where $X$ varies over a range, say $\pm L$, and provides a measure for the degree of anisotropic acceleration. For example, if $X=0$, we simply recover $d(\Lambda)$, and if $X=-1$, we are left with $d(\Lambda=0)$. Introducing a unit vector $\hat{l}_{i}$ that points in the direction of the $i^{\rm th}$ SN, and a vector $\vec{L}$, that points in the direction of `preferred' acceleration, we can suitably choose $X=\hat{l}_{i}\cdot\vec{L}$, thus Eq.~3 becomes

\begin{equation}
\delta_{M,i} = D(z_{i})(1+\hat{l}_{i}\cdot\vec{L}) \label{eqn:dpmodel}
\end{equation}
with,
\begin{equation}
\delta_{M,i} = \frac{d(\Lambda_{A})-d(\Lambda=0)}{d(\Lambda=0)} \label{eqn:ML02b}
\end{equation}
\begin{equation}
D(z_{i}) = \frac{d(\Lambda)-d(\Lambda=0)}{d(\Lambda=0)} \label{eqn:ML02c}
\end{equation}
where $D(z_{i})$ is the `deviation parameter' for the $i^{\rm th}$ SN with redshift $z_{i}$. Using this formalism,
we can introduce the measured quantity $\delta_i$ (recall that $d_{SN}$ is the measured luminosity distance to a given SN),

\begin{equation}
\delta_{i} = \frac{d_{SN}-d(\Lambda=0)}{d(\Lambda=0)} \label{eqn:ML02d}
\end{equation}

Thus, the vector $\vec{L}$ provides an estimate for the magnitude and direction of the dipole, whilst the deviation parameter describes how the dipole changes with redshift. If a dipole in dark energy were to exist, we would expect it to be more exposed at higher redshifts where the acceleration due to dark energy is more apparent. We have therefore opted for a deviation parameter, and hence dipole, that increases with redshift. Ideally, one would prefer to divide the data into `redshift shells' to derive an arbitrary form for the dipole, however, we are at present limited by the relatively small number of SNe. By choosing the model in this way, we are testing for the significance of a dipole in the acceleration term in the SNe data. A dipole with a different $z$ dependence would still be detected by our analysis but the significance level would not then be correctly estimated.

\subsubsection{The Likelihood Function}
The dipolar model (Eq.~\ref{eqn:dpmodel}) forms the basis for the present analysis, and is employed using a maximum likelihood strategy. As the observations are not \emph{uniformly} distributed around the sky, we pose the question, given the directions $\hat{l}_{i}$ of the observed SNe, what is the most probable vector $\vec{L}$ that gives an anisotropy of the form in Eq.~\ref{eqn:dpmodel}? Non-uniform sky coverage will therefore only result in a larger error ellipsoid for $\vec{L}$.

Within this framework, poorly measured SNe that deviate substantially from the concordance model will produce spurious results. We therefore weight each SN according to its observational error. Suppose the measurements $\delta_{i}$ come from true values $\Delta_{i}$ scattered by observational errors $\sigma_{i}$. Then, given an observation $\delta_{i}$, the probability of it arising from a true value $\Delta_{i}$ is
\begin{equation}
Pr(\delta_{i}|\Delta_{i})=\frac{1}{\sqrt{2\pi}\sigma_{i}}\exp\bigg(\frac{-(\delta_{i}-\Delta_{i})^{2}}{2\sigma_{i}^{2}}\bigg)
\end{equation}
where $\sigma_{i}$ is the observers error estimate for the $i^{\rm th}$ SN. Similarly, the probability that the true $\Delta_{i}$ arise from a scatter $\sigma$ about the proposed dipolar model (Eq.~\ref{eqn:dpmodel}), is
\begin{equation}
Pr(\Delta_{i}|\delta_{M,i})=\frac{1}{\sqrt{2\pi}\sigma}\exp\bigg(\frac{-(\Delta_{i}-\delta_{M,i})^{2}}{2\sigma^{2}}\bigg)
\end{equation}
The probability, therefore, of measuring the $\delta_{i}$ given our model is found by integrating over all true values
\begin{eqnarray}
\lefteqn{ Pr(\delta_{i}|\delta_{M,i})=\int_{-\infty}^{\infty} Pr(\delta_{i}|\Delta_{i}) \,\, Pr(\Delta_{i}|\delta_{M,i}) \,\, {\rm d}\Delta_{i} } \nonumber
\\
\lefteqn{ Pr(\delta_{i}|\delta_{M,i})=\frac{1}{\sqrt{2\pi(\sigma^{2}+\sigma_{i}^{2})}}\exp\bigg(\frac{-(\delta_{i}-\delta_{M,i})^{2}}{2(\sigma^{2}+\sigma_{i}^{2})}\bigg) \label{eqn:prob_data_given_model} }
\end{eqnarray}
The log-likelihood function is then given by
\begin{equation}
\mathcal{L}=\ln\bigg[ \prod_{i} Pr(\delta_{i}|\delta_{M,i}) \bigg]\label{eqn:likelihoodfunc}
\end{equation}

\subsubsection{Maximum Likelihood}\label{sec:max_LLH}
Substitution of the dipolar model into Eq.~\ref{eqn:likelihoodfunc}, and maximising the likelihood over all choices of the model parameters $\vec{L}$ and $\sigma$, $\partial\mathcal{L}/\partial\vec{L}\,=\,\vec{0}$ yields an equation of the form
\begin{equation}
\mathbf{A} \cdot \vec{L} = \vec{V}
\label{eqn:marg_l}
\end{equation}
where $\mathbf{A}$ (a rank 2 tensor), and $\vec{V}$ are given by
\begin{equation}
\mathbf{A} = \sum_{i} \frac{D(z_{i})^{2}}{\sigma^{2}+\sigma_{i}^{2}}\,\hat{l}_{i}\,\hat{l}_{i}
\label{eqn:marg_A}
\end{equation}
\begin{equation}
\vec{V} = \sum_{i}\frac{(\delta_{i}-\,D(z_{i}))\,D(z_{i})}{\sigma^{2}+\sigma_{i}^{2}}\,\hat{l}_{i}
\label{eqn:marg_V}
\end{equation}
$\partial\mathcal{L}/\partial\sigma\,=\,0$ gives
\begin{equation}
\sum_{i}\frac{1}{\sigma^{2}+\sigma_{i}^{2}}=\sum_{i}\frac{(\delta_{i}-\delta_{M,i})^{2}}{(\sigma^{2}+\sigma_{i}^{2})^{2}}
\label{eqn:marg_sigma}
\end{equation}

We must now simultaneously solve the coupled non-linear Eqs.~\ref{eqn:marg_l} and \ref{eqn:marg_sigma}. The most efficient pathway to a solution is to construct a set of $\sigma$ values. Using Eq.~\ref{eqn:marg_l}, one can invert $\mathbf{A}$, and find $\vec{L}$ for each $\sigma$. We then use Eq.~\ref{eqn:marg_sigma} to solve for the $\sigma$ value that maximises $\mathcal{L}$.

\begin{table*}\label{tab:results}
\centering
\begin{minipage}[c]{0.99\textwidth}
\centering
     \textbf{Table 1.} \textsc{Parameter Fitting Results}\\
    \begin{tabular}{lcccccccc}
    \hline
    \hline
  \multicolumn{1}{c}{Model}
& \multicolumn{1}{c}{$|\vec{L}|\times10^{2}$} 
& \multicolumn{1}{c}{$l^{\dag}$}
& \multicolumn{1}{c}{$b^{\dag}$}
& \multicolumn{1}{c}{$\sigma\times10^{2}$}
& \multicolumn{1}{c}{$|\vec{L}_{M}|\times10^{2}$}
& \multicolumn{1}{c}{$\mathcal{L}^{\ddag}$}
& \multicolumn{1}{c}{$L_{err}/|\vec{L}|$}
& \multicolumn{1}{c}{$Pr(\vec{L})$}
\\
  \hline
$^{\rm a}$Weighted Dipolar Model ($z>0.2$)      & 7.14 & 287$^{\circ}$ & 46$^{\circ}$ & 0.00 & 15.7 & 105.5 & 1.04 & $18\,\%$  ($0.23\sigma$) \\
$^{\rm b}$Weighted Dipolar Model ($0.2<z<0.56$) & 3.90 & 258$^{\circ}$ &  9$^{\circ}$ & 0.00 & 19.1 & 82.9 & 3.44 & $0.6\,\%$ ($0.008\sigma$) \\
$^{\rm c}$Weighted Dipolar Model ($z>0.56$)     & 13.8 & 309$^{\circ}$ & 43$^{\circ}$ & 3.52 & 26.1 & 23.3 & 0.83 & $31\,\%$  ($0.39\sigma$) \\
$^{\rm d}$Higher Quality List                   & 10.0 & 118$^{\circ}$ & 25$^{\circ}$ & 14.2 & 20.8 & 97.1 & 1.03 & $18\,\%$  ($0.23\sigma$) \\
    \hline
    \end{tabular}
    \smallskip
\\
\end{minipage}
$^{\rm a,b,c,d}$ These models are referred to in the text as Sample 1,2,3,4. $^{\dag}$Galactic Coordinates. $^{\ddag}$Given by Eq.~\ref{eqn:likelihoodfunc}.
\end{table*}

\subsubsection{Other Solution Pathways}\label{sec:unweighted}
The non-linearity of the above equations is introduced through the weighting of each SN by its associated error. If we were to weight all SNe equally (i.e. $\sigma_{i}=constant$ for all $i$), Eqs.~\ref{eqn:marg_A}-\ref{eqn:marg_sigma} then take the form

\begin{eqnarray}
\lefteqn{ \mathbf{A} = \sum_{i} D(z_{i})^{2}\,\hat{l}_{i}\,\hat{l}_{i} }
\label{eqn:marg_A2}
\\
\lefteqn{ \vec{V} = \sum_{i}(\delta_{i}-\,D(z_{i}))\,D(z_{i})\,\hat{l}_{i} }
\label{eqn:marg_V2}
\\
\lefteqn{ \sigma^{2}=\frac{1}{N}\sum_{i}(\delta_{i}-\delta_{M,i})^{2} }
\label{eqn:marg_sigma2}
\end{eqnarray}
where $N$ is the total number of SNe in the sample. We can now invert $\mathbf{A}$ and uniquely solve for $\vec{L}$, and then use Eq.~\ref{eqn:marg_sigma2} to solve for the $\sigma$ that maximises the likelihood, $\mathcal{L}$.

\section{Results and Discussion}
In this section we present the results from our maximum likelihood analysis, and introduce a method to quantify the significance of our derived parameter values. Before proceeding, it is noteworthy that the adopted maximum likelihood strategy detailed in \S\ref{sec:max_like_strategy} requires all SNe coordinates to be converted from Galactic to Cartesian coordinates\footnote{The $x$ and $z$ Cartesian axes are respectively directed towards the Galactic centre and North pole.}.

\subsection{Significance of the Dipolar Models}\label{sec:significance}
We ask ourselves, what is the probability that the vector $\vec{L}$, that maximises the likelihood, is true? The covariance matrix for $\vec{L}$ is given by the inverse of Eq.~\ref{eqn:marg_A} (i.e. $\mathbf{A}^{-1}$), or, for the solution pathway outlined in \S\ref{sec:unweighted}, $\sigma^{2}\mathbf{A}^{-1}$. Given the eigenvalues ($\lambda_{1},\lambda_{2},\lambda_{3}$) and corresponding eigenvectors ($\hat{e}_{1}$,\,$\hat{e}_{2}$,\,$\hat{e}_{3}$) of the covariance matrix, the three semi-principal axes of the error ellipsoid for $\vec{L}$ are $\vec{s}_{1,2,3}=\sqrt{\lambda_{1,2,3}}\,\,\hat{e}_{1,2,3}$. The density of probability associated with the vector $\vec{L}$ is therefore of the form
\begin{equation}
p(\vec{L}) = \frac{1}{\sqrt{(2\pi)^{3}\,\lambda_{1}\,\lambda_{2}\,\lambda_{3}}}\exp-\Big(\frac{e_{1}^{2}}{2\lambda_{1}}+\frac{e_{2}^{2}}{2\lambda_{2}}+\frac{e_{3}^{2}}{2\lambda_{3}}\Big)
\end{equation}
where $e_{1}$, $e_{2}$ and $e_{3}$ are the Cartesian coordinate axes in the direction of the principal axes of the error ellipsoid. The length of the vector $\vec{L}$ enclosed by the error ellipsoid is given by
\begin{equation}
L_{err}=[(\hat{L}\cdot\vec{s}_{1})^{2}+(\hat{L}\cdot\vec{s}_{2})^{2}+(\hat{L}\cdot\vec{s}_{3})^{2}]^{1/2}
\end{equation}
We can now introduce the quantity $\mu=L_{err}/|\vec{L}|$, which defines the fraction of $\vec{L}$ enclosed by the error ellipsoid. Suppose we now expand (or contract) the error ellipsoid such that it contains a volume $V_{E}$, that \emph{just} encompasses $\vec{L}$. The probability that our derived $\vec{L}$ is true, is therefore given by
\begin{equation}
Pr(\vec{L})=\int_{V_{E}}p(\vec{L})\,{\rm d}V_{E}
\end{equation}
which after converting to spherical coordinates yields
\begin{eqnarray}
\lefteqn{ Pr(\vec{L})=\frac{4\pi}{(\pi)^{3/2}}\int_{0}^{1/\mu\sqrt{2}}u^{2}\exp\big({-u^{2}}\big)\,{\rm d}u } \nonumber
\\
\nonumber
\\
\lefteqn{ Pr(\vec{L})=\rm{erf}(1/\mu\sqrt{2}) - \sqrt{\frac{2}{\mu^{2}\pi}}\exp\big(-1/2\mu^{2}\big) }
\label{eqn:prob_linside}
\end{eqnarray}

\begin{figure}
\centering
\includegraphics[width=7.8cm]{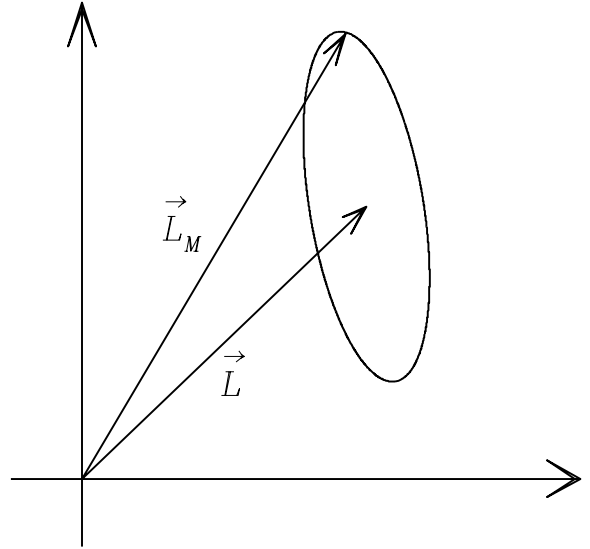}
\caption{
An illustration of $\vec{L}$ and its error ellipsoid. The largest vector contained within $\vec{L}$'s error ellipsoid, $\vec{L}_{M}$, is also shown.
}\label{fig:lm}
\end{figure}

Additionally, one can find a vector, $\vec{L}_{M}$, that is the largest vector contained within the error ellipsoid of $\vec{L}$ (see Fig.~\ref{fig:lm}). If we are \textit{certain} that a dipole exists, and that $\vec{L}$ is within its error ellipsoid, the quantity $|\vec{L}_{M}|$ is then a measure of the maximum allowed increase in the acceleration, irrespective of the direction of $\vec{L}$.

\subsection{The Weighted Dipolar Model}

\begin{figure}
\includegraphics[width=8cm]{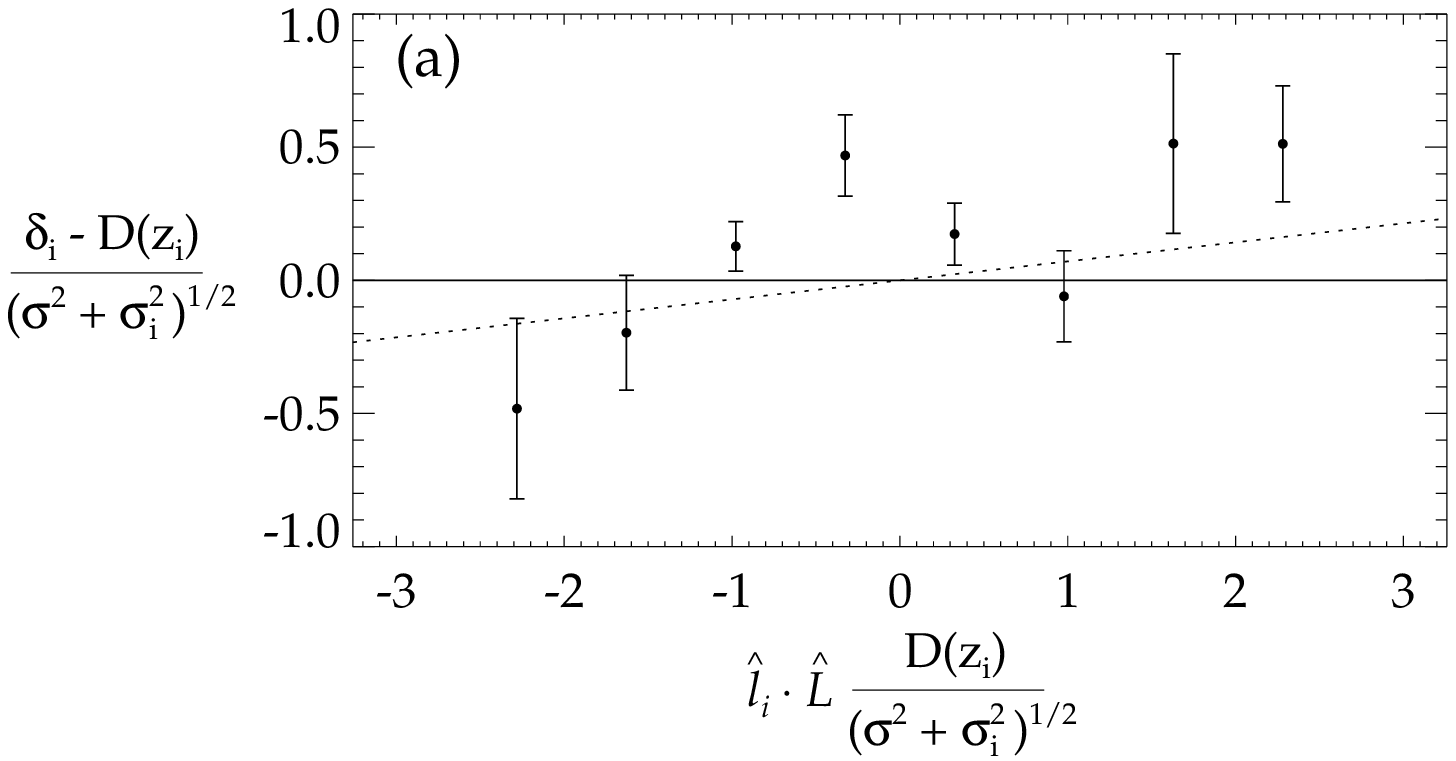}
\newline\newline
\includegraphics[width=8cm]{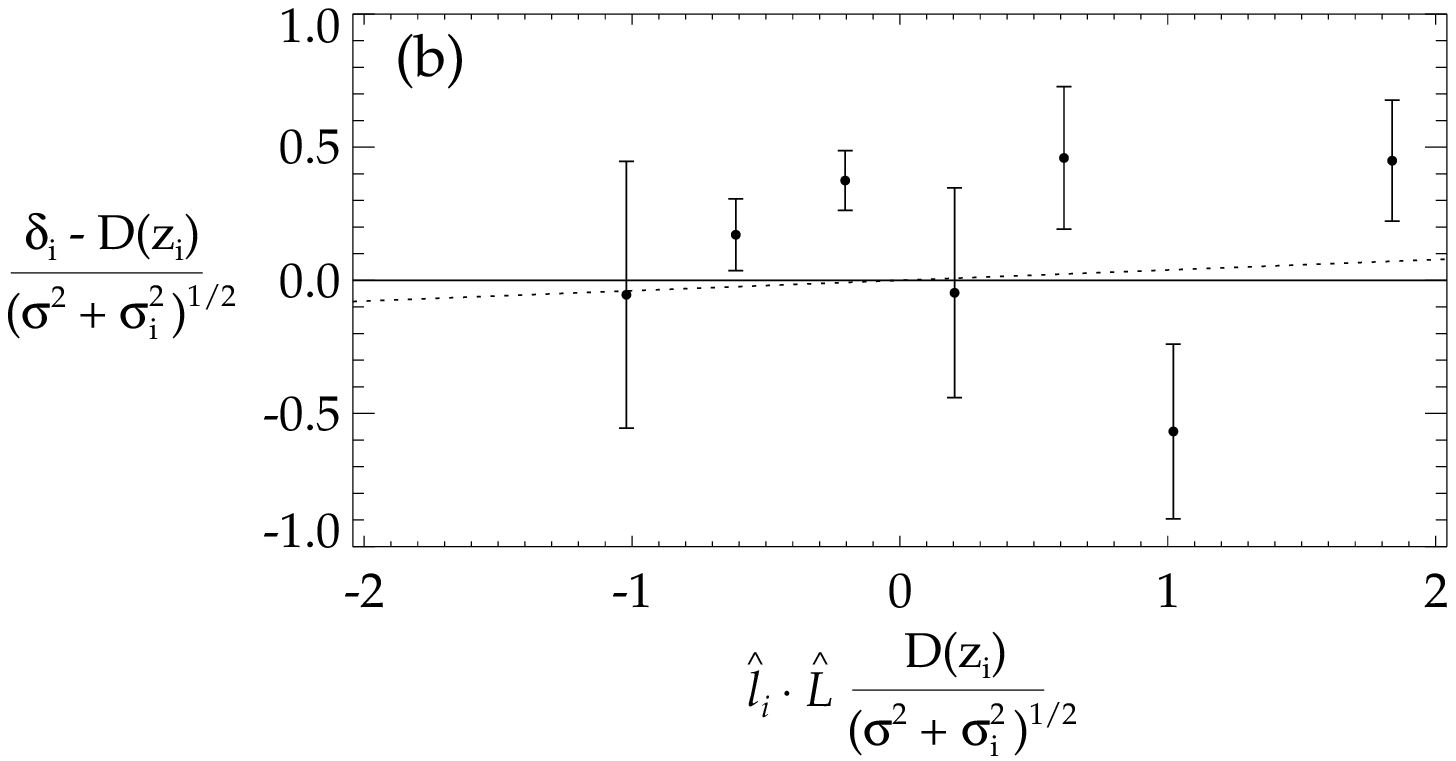}
\newline\newline
\includegraphics[width=8cm]{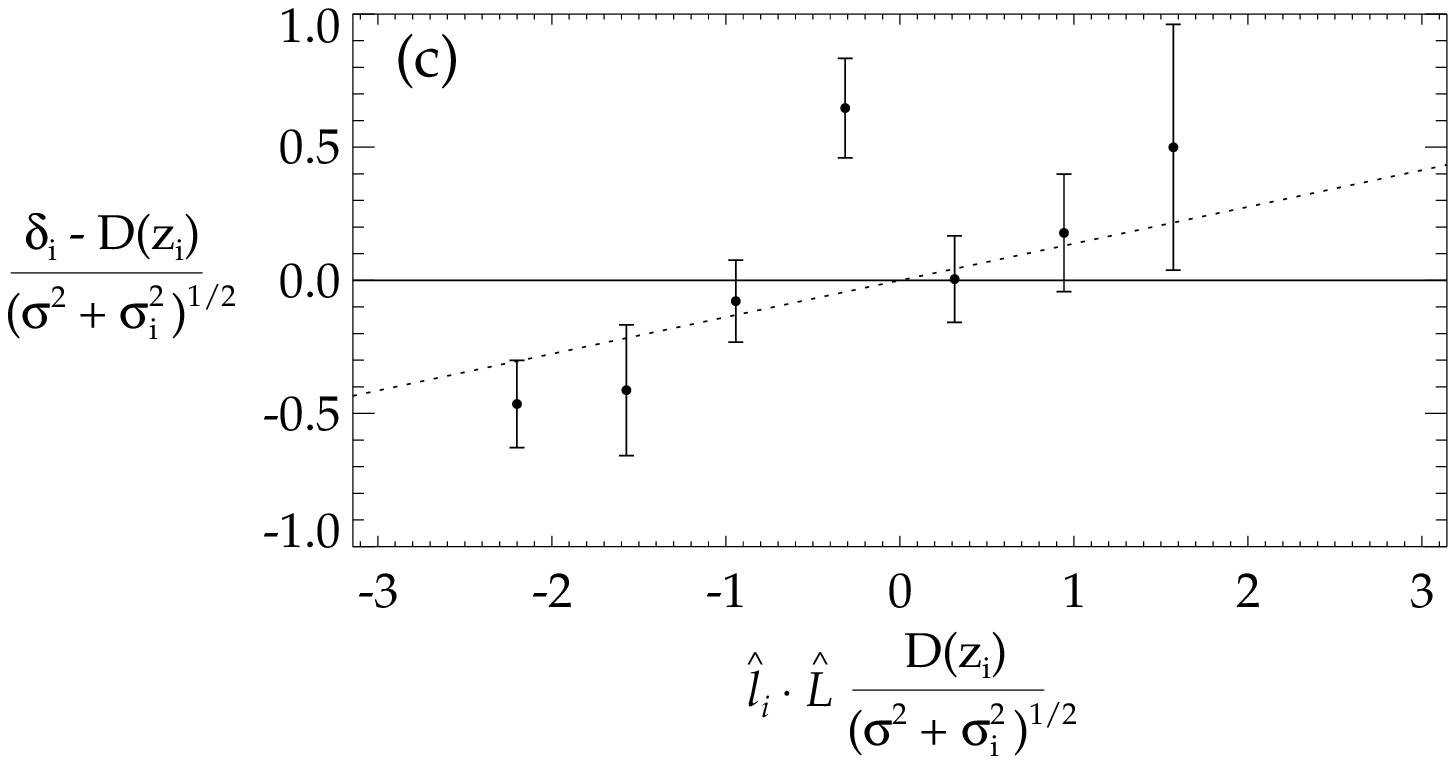}
\newline\newline
\includegraphics[width=8cm]{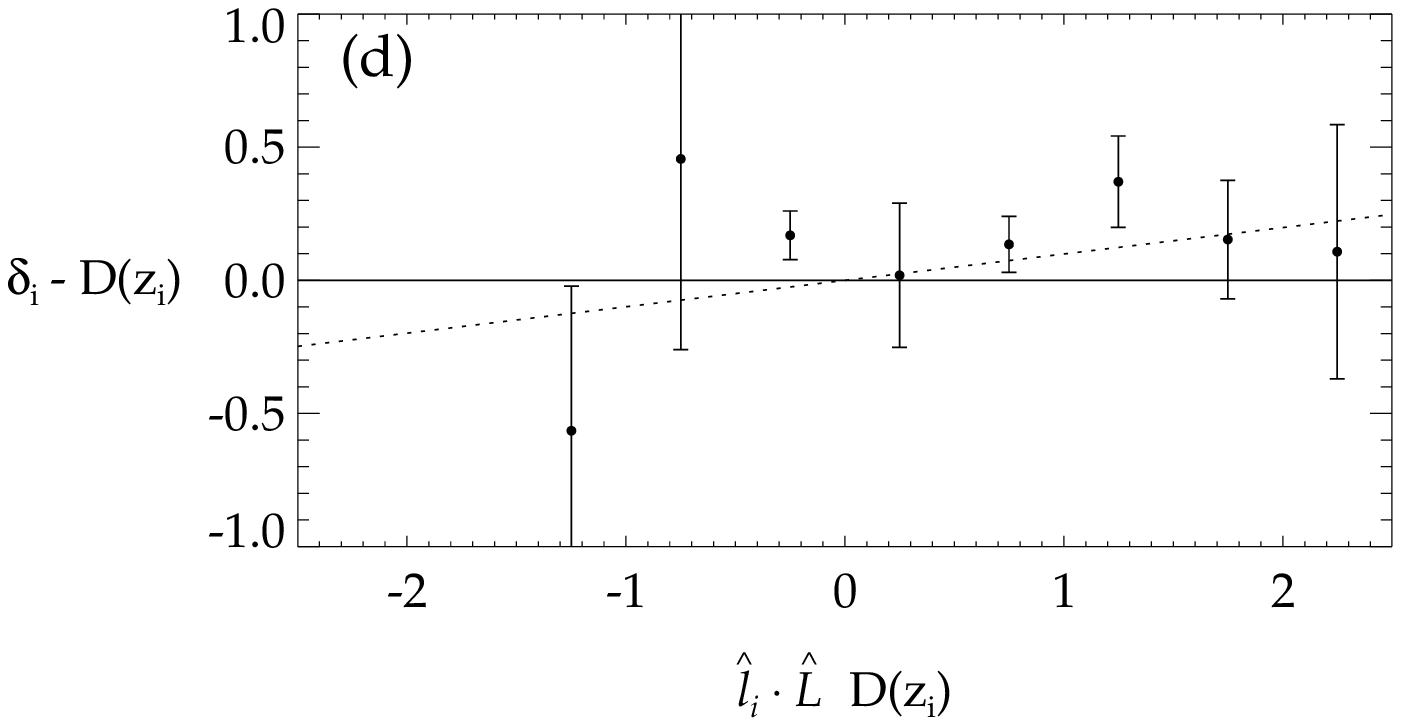}
\caption{
For all figures, the y-axis corresponds to the case where dark energy is isotropic (i.e. $X=\hat{l}_{i}\cdot\vec{L}=0$), and the x-axis corresponds to the cosine of the angle between the $i^{\rm th}$ SN and the `preferred' direction $\vec{L}$, weighted by the deviation parameter and the individual errors, so that poorly measured SNe contribute less, as required by the dipolar model. Thus, the dotted line has a gradient of $|\vec{L}|$. For a perfect fit to the dipolar model, the data should lie along this dotted line. The data are averaged in 10 equally spaced bins, where the errors represent the standard error in the mean of the given bin. The domain is set to be the maximum abscissa of the unbinned data.
(a) Weighted dipolar model (Sample 1) $z>0.2$. 
(b) Weighted dipolar model (Sample 2) $0.2<z<0.56$. 
(c) Weighted dipolar model (Sample 3) $z>0.56$.
(d) Higher Quality List (Sample 4).
}
\label{fig:liL}
\end{figure}

\begin{figure}
\includegraphics[width=8cm]{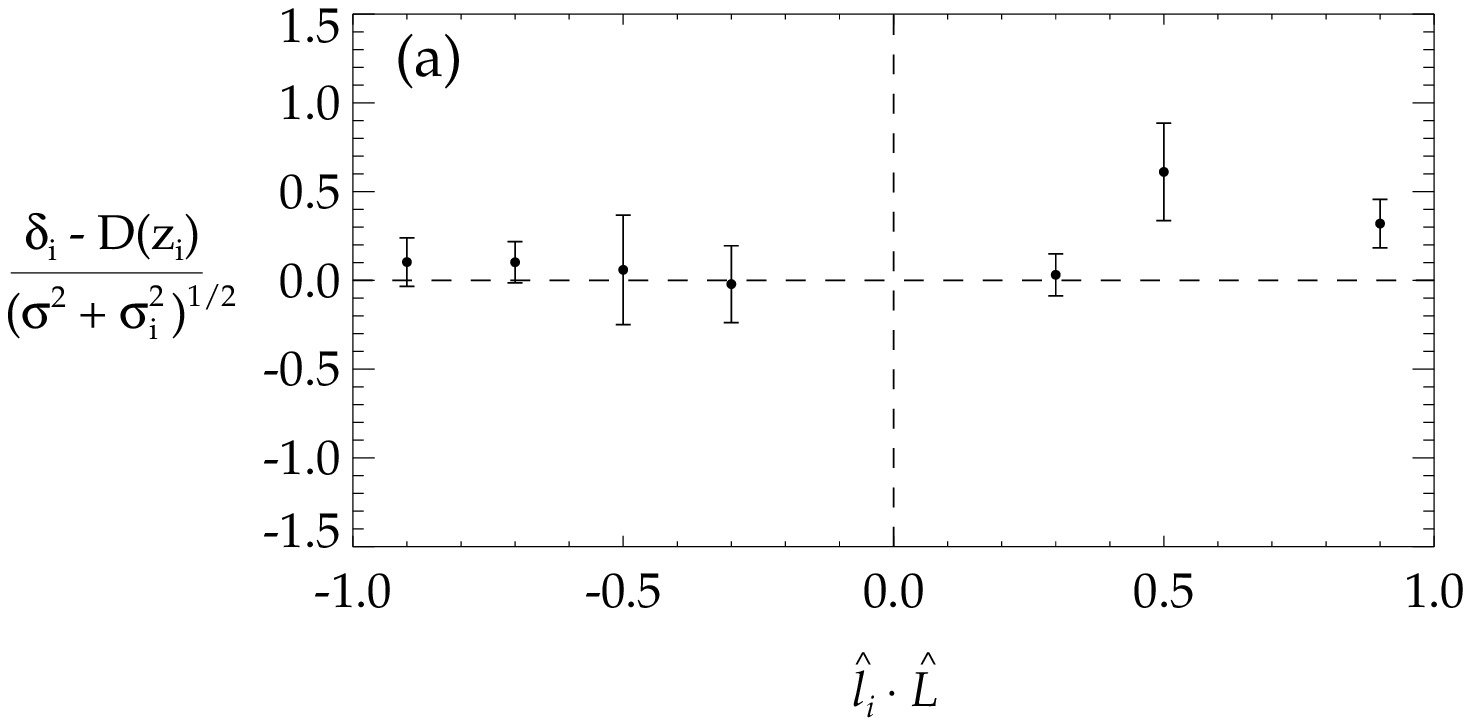}
\newline\newline
\includegraphics[width=8cm]{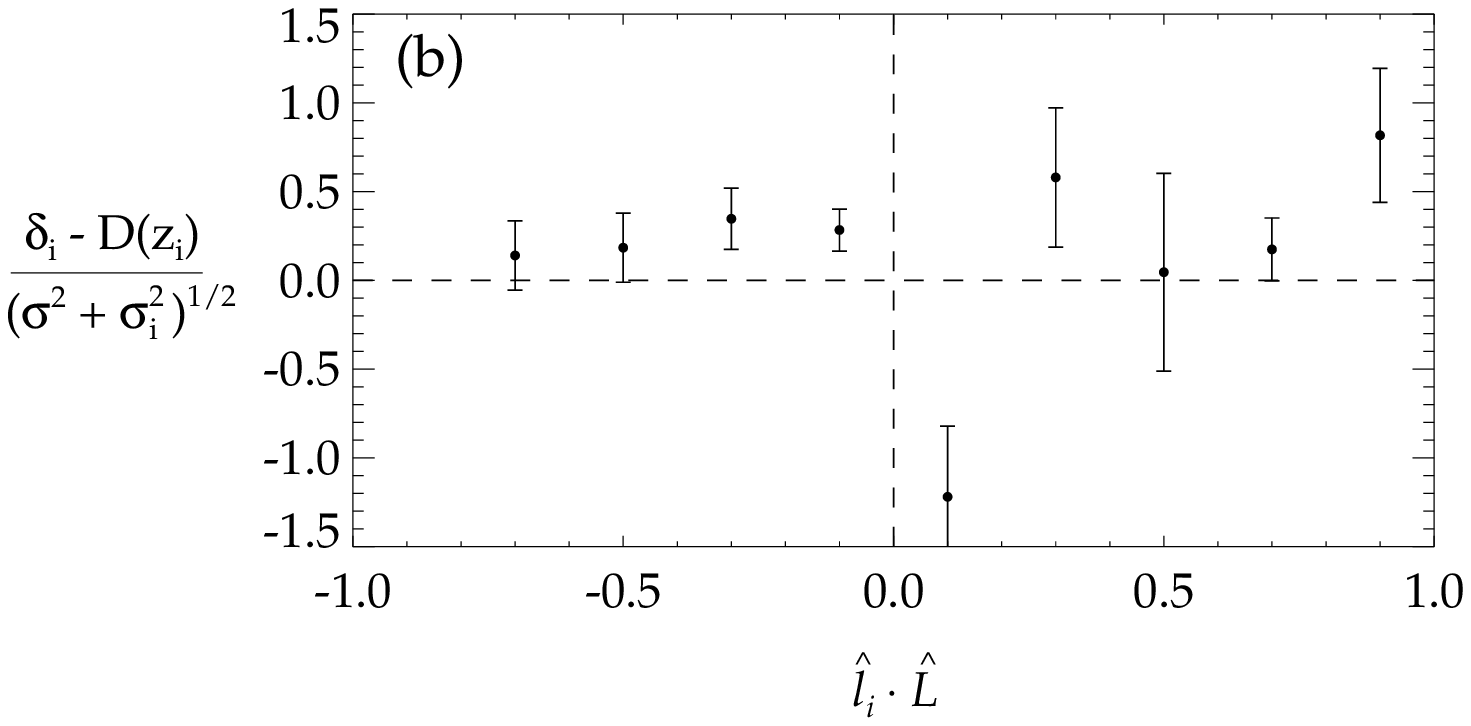}
\newline\newline
\includegraphics[width=8cm]{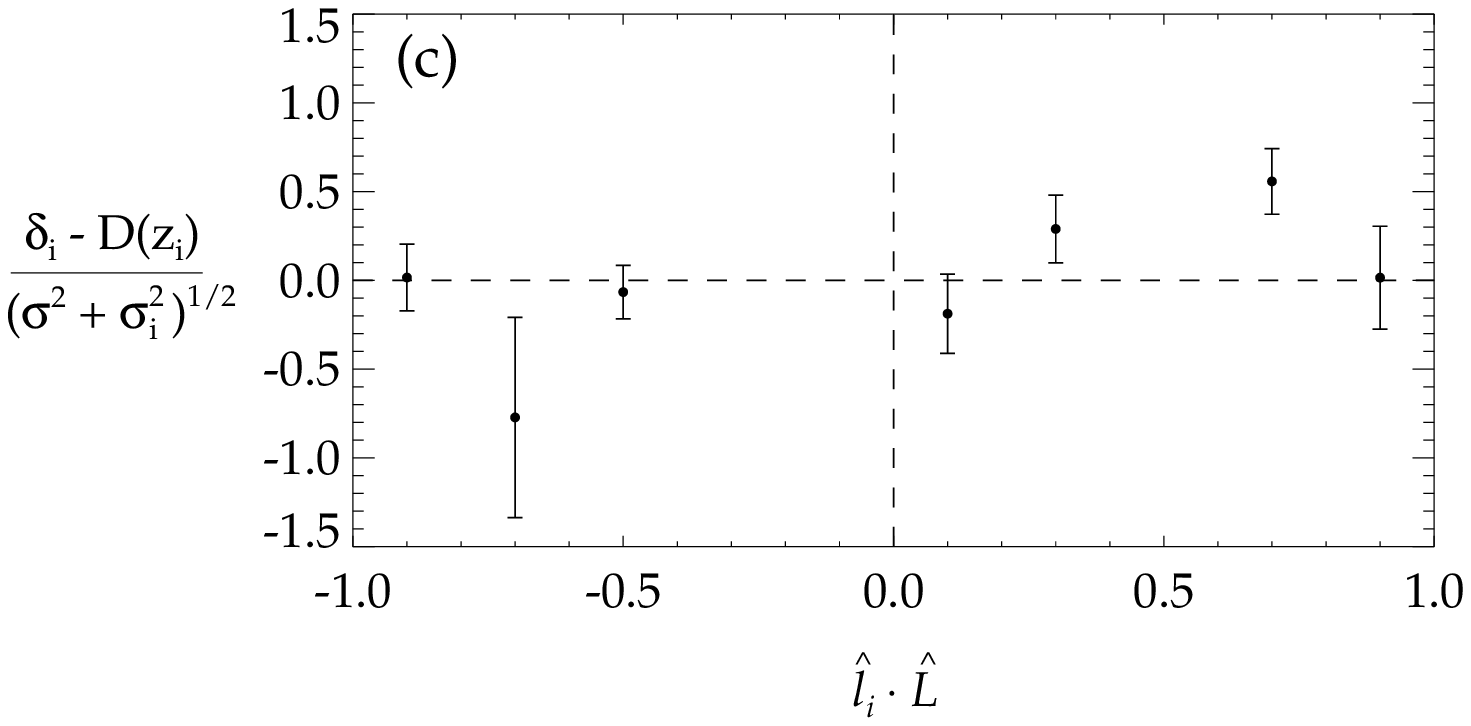}
\newline\newline
\includegraphics[width=8cm]{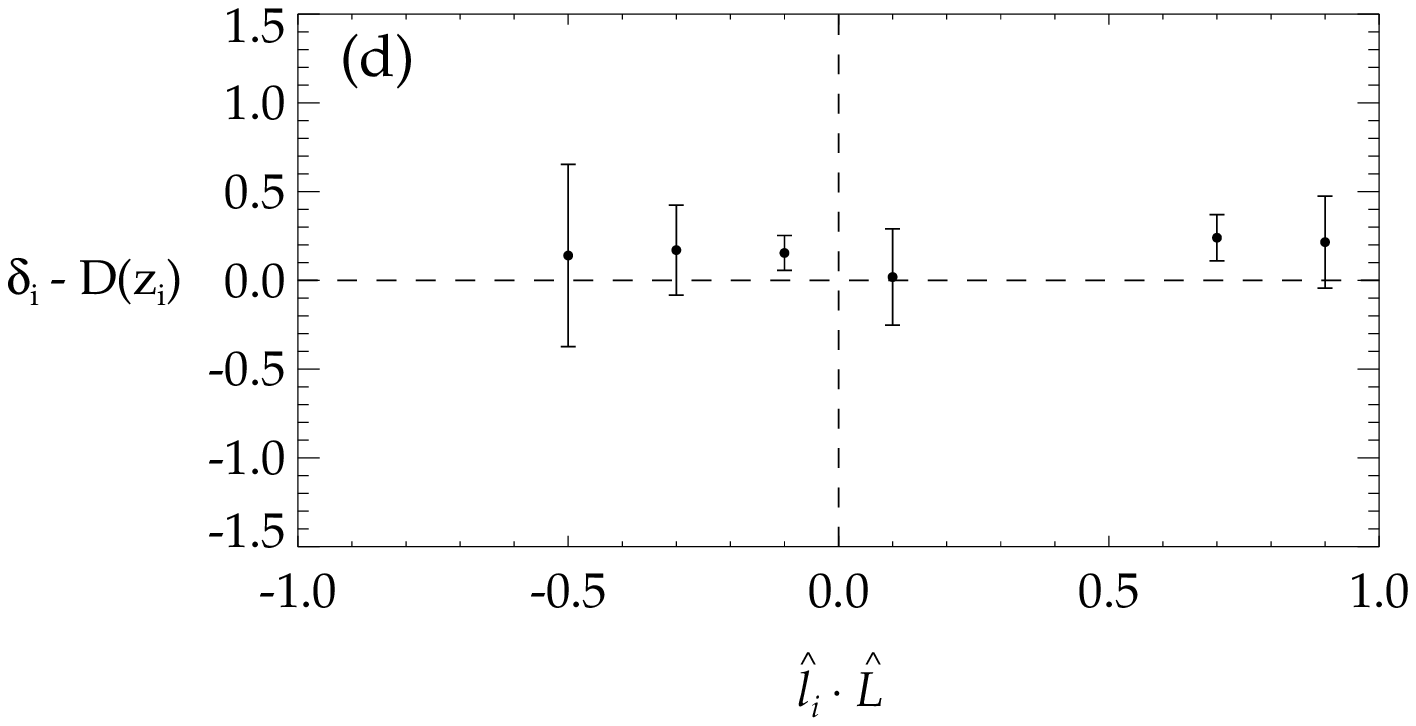}
\caption{
For all figures, the y-axis again corresponds to the case where dark energy is isotropic (i.e. the same as Fig.~\ref{fig:liL}), however the x-axis now simply corresponds to the cosine of the angle between the $i^{\rm th}$ SN and the `preferred' direction $\hat{L}$. Thus, we have used the derived $\hat{L}$ for each Sample, to determine whether or not there are any departures from isotropy in the weighted data. A general trend from bottom-left to top-right would be expected if a preferred direction did exist. The data are averaged in 10 equally spaced bins, where the errors represent the standard error in the mean of the given bin.
(a) Weighted dipolar model (Sample 1) $z>0.2$. 
(b) Weighted dipolar model (Sample 2) $0.2<z<0.56$. 
(c) Weighted dipolar model (Sample 3) $z>0.56$.
(d) Higher Quality List (Sample 4).
}
\label{fig:liLtwo}
\end{figure}

We now consider the weighted dipolar model; weighting all SNe according to their measured errors, and implement the solution strategy outlined in \S\,\ref{sec:max_LLH}. The model parameters that maximise the likelihood function are presented in Table~1 (Sample 1), where the components of $\vec{L}$ are given in Galactic coordinates. For this case, $\sigma^2$ is negligible compared to the minimum $\sigma_{i}^{2}$ in the sample (refer to Eq.~\ref{eqn:prob_data_given_model}). The fitting results are also illustrated in Fig.~\ref{fig:liL}(a), where the data are averaged in 10 equally spaced bins, and the error bars correspond to the standard error in the mean of each bin. Bins containing $\leq2$ points are not shown.

A striking observation of the results from the weighted dipolar model, is the similarity in the direction of $\vec{L}$ to the CMB dipole: $(l,b) = (263.99^{\circ} \pm 0.14^{\circ}, 48.26^{\circ} \pm 0.03^{\circ})$ \citep{hin09}. Although there is a relatively minor significance, this coincidence is certainly worth further scrutiny.

To investigate whether the higher or lower redshift SNe, or both, favour this direction, we divide the SNe into two subsamples, each containing 125 SNe, and then run the weighted dipolar model routine on the two subsamples. This division corresponds to a redshift $z=0.56$. The results are again presented in Table~1, and illustrated in Fig.~\ref{fig:liL}(b) and \ref{fig:liL}(c) for the two subsamples $0.2<z<0.56$ (Sample 2) and $z>0.56$ (Sample 3) respectively. As can be appreciated from Fig.~\ref{fig:liL}(b), and the final column of Table~1, the lower redshift subsample does not resemble the anisotropic model at all. The higher redshift subsample presented in Fig.~\ref{fig:liL}(c), on the other hand, seems more consistent with the results of Sample 1.

In order to test whether the Union compilation exhibit a dipolar acceleration, we plot the weighted data as a function of $\hat{l}_{i}\cdot\hat{L}$ (i.e. the cosine of the angle between the $i^{\rm th}$ SN and the `preferred direction') in Fig.~\ref{fig:liLtwo}(a)-(c). Should a preferred direction exist, one would expect the data to show a general trend from the bottom-left quadrant to the top-right quadrant, a feature that is not observed in these samples. One could tentatively claim that this is true for Sample 3 (in Fig.~\ref{fig:liLtwo}(c)), but the significance is very low.

\subsection{The Higher Quality List}

\begin{figure}
\includegraphics[width=8cm]{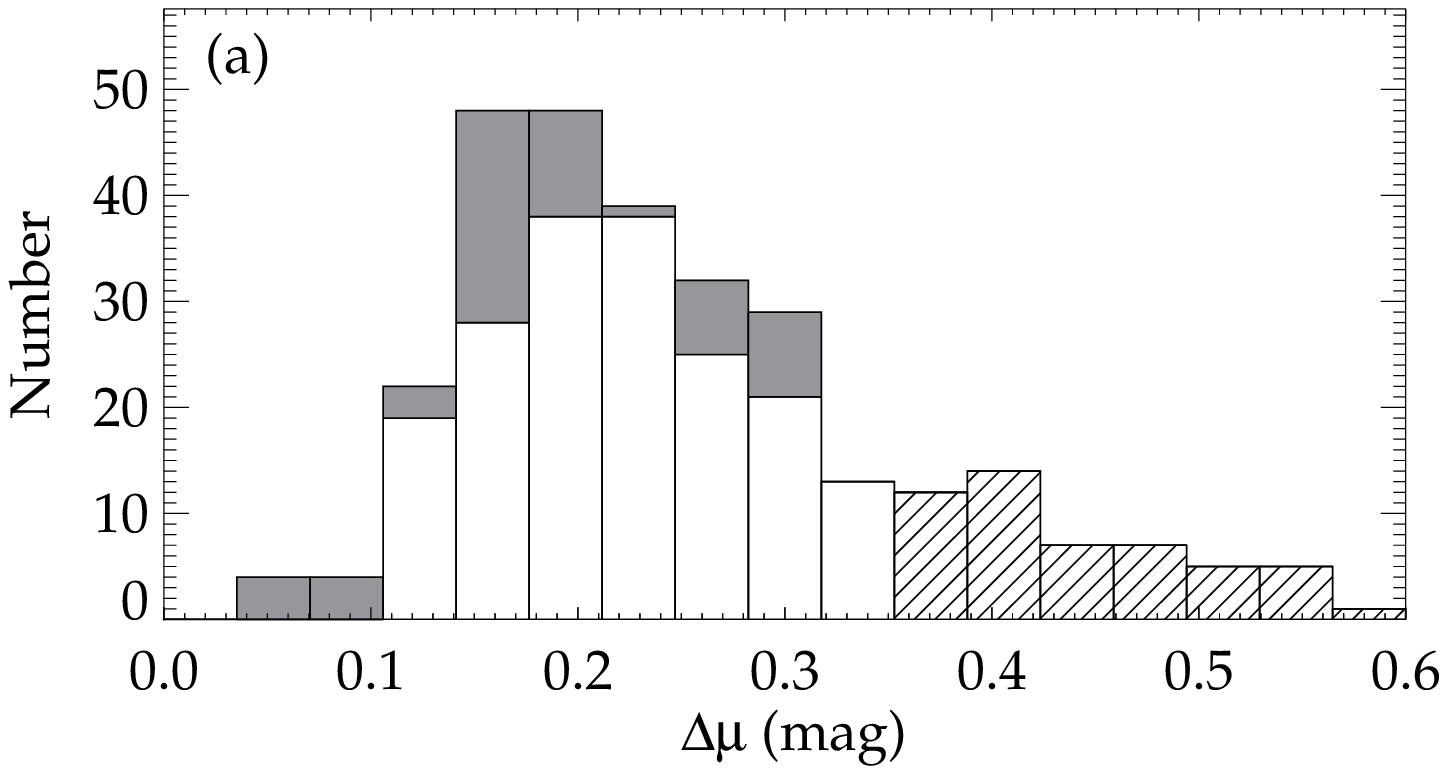}
\newline\newline
\includegraphics[width=8cm]{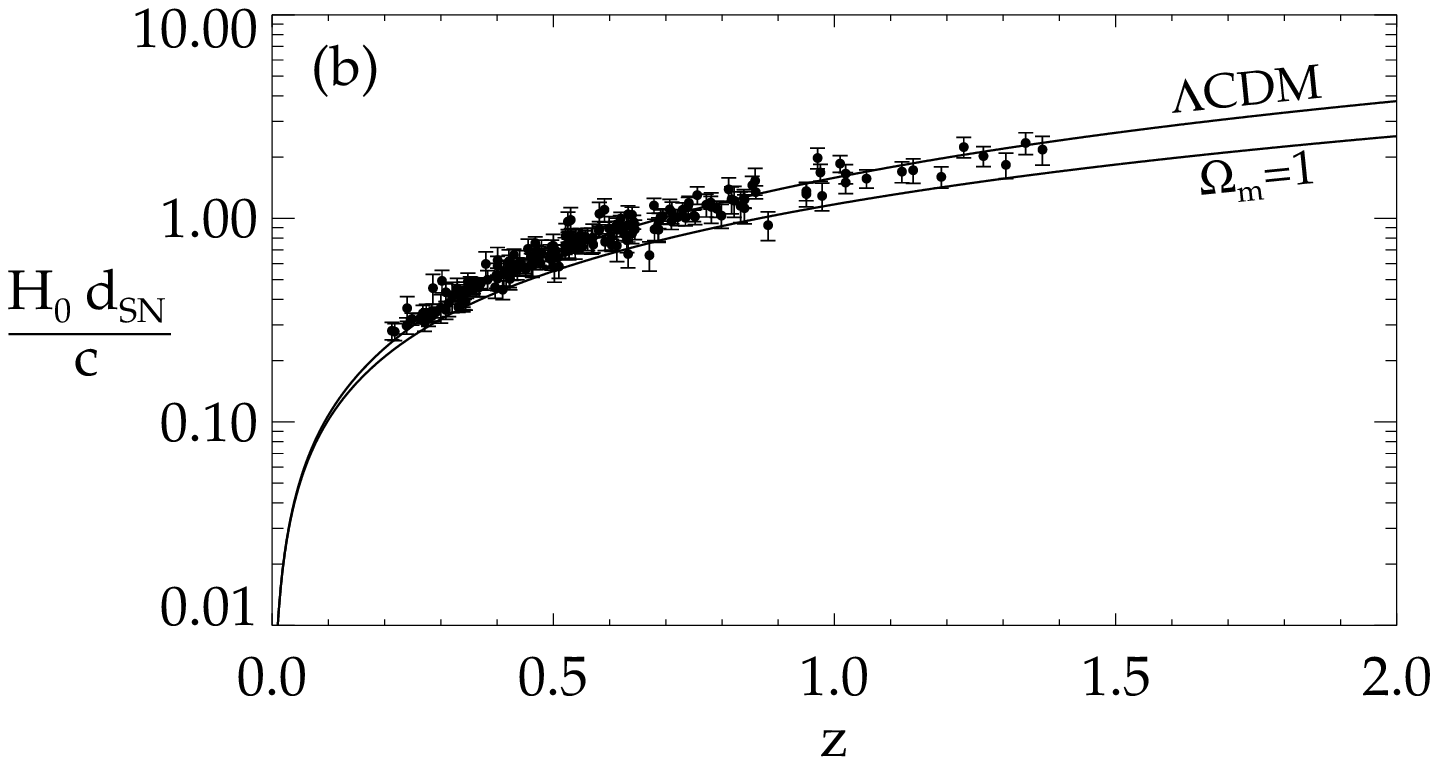}
\caption{
(a) Distribution of the SNe errors from the Union compilation. The light grey shaded regions correspond to the SNe with $z<0.2$. The hatched regions correspond to the SNe that have $\Delta\mu\ge0.35$. The unshaded histogram represents the Higher Quality List, with $z>0.2$ and $\Delta\mu<0.35$.
(b) A Hubble plot of the Higher Quality List (symbols) and Hubble law for $d(\Lambda)$ (upper solid line) and $d(\Lambda=0)$ (lower solid line).}
\label{fig:hq_hist}
\end{figure}

In \S~\ref{sec:unweighted}, an alternative linear method was described to solve for the model parameters, however, as mentioned previously, one must be cautious when assigning each SN measurement an equal weighting, as poorly measured SNe that deviate substantially from the Hubble law could produce spurious results. These concerns should be alleviated by compiling a `Higher Quality List', containing only the best quality SNe (Sample 4). We therefore restrict the sample to contain only the best measured SNe; SNe with uncertainties in magnitude $>0.35$ are rejected. The distribution of accepted and rejected SNe are presented in Fig.~\ref{fig:hq_hist}(a). The Hubble diagram for the Higher Quality List is shown in Fig.~\ref{fig:hq_hist}(b).

One would expect the results of the weighted dipolar model (Sample 1) to be consistent with the Higher Quality List, however, this does not seem to be the case. The direction of $\vec{L}$ has changed by $108^{\circ}$, and therefore does not lend support to the results of Sample 1. For completeness, we also illustrate these results in Fig.~\ref{fig:liL}(d), and Fig.~\ref{fig:liLtwo}(d), however, neither demonstrate the trend that would be expected if an anisotropy existed in the Union SNe compilation.

\section{Conclusions}

The acceleration of the Universe appears to be equal in all directions. The anisotropy we find is a $14\,\%\,\pm\,12\,\%$ increase toward $(l,b)=(309^{\circ},43^{\circ})$ and a corresponding decrease in the opposite direction $(l,b)=(129^{\circ},-43^{\circ})$, but the effect is only apparent in the higher redshift group with $z>0.56$. These are those that exhibit the largest net acceleration. The direction of this weak anisotropy is only $31^{\circ}$ from that of the CMB dipole as seen from the Sun, $(l,b)=(264^{\circ},48.3^{\circ})$, or $17^{\circ}$ from that of the dipole observed in the CMB frame of the elliptical galaxies sample with $v<2000\,{\rm km\, s}^{-1}$, $(l,b)=(311^{\circ},26^{\circ})$ \citep{lyn88}. The Union SN redshifts are corrected to the CMB system of rest, so their zero point cannot cause this effect.

It is noteworthy that both \citet{kol01} and \citet{gup08} find a directional dependent systematic in the SNe data of similar significance to what we have found here, although, the directions are somewhat different. The present dataset cannot exclude a $26\,\%$ (i.e. 14+12) increase in the acceleration toward $(l,b)=(309^{\circ},43^{\circ})$ and together with a $26\,\%$ decrease in the opposite direction. More high accuracy $z>0.5$ data from directions within $45^{\circ}$ of $(l,b)=(309^{\circ},43^{\circ})$ or $(l,b)=(129^{\circ},-43^{\circ})$ are needed to eliminate any possible anisotropy.

\section*{Acknowledgements}

We wish to thank the referee for their valuable comments that helped to improve this article. RC thanks D. Quinn for useful discussions. RC is jointly funded by the Cambridge Overseas Trust and the Cambridge Commonwealth/Australia Trust with an Allen Cambridge Australia Trust Scholarship.

\bsp

\label{lastpage}


\begin{thebibliography}{99}
\bibitem[\protect\citeauthoryear{Alnes, Amarzguioui \& Gr\o n}{Alnes \etal}{2006}]{aln06}
Alnes H., Amarzguioui M., Gr\o n \O., 2006, Phys. Rev. D, 73, 083519
\bibitem[\protect\citeauthoryear{Battye \& Moss}{Battye \& Moss}{2006}]{bat06}
Battye R.~A., Moss A., 2006, Phys. Rev. D, 74, 041301
\bibitem[\protect\citeauthoryear{Cooray, Holz \& Caldwell}{Cooray \etal}{2008}]{coo08}
Cooray A., Holz D.~E., Caldwell R., 2008, arXiv: 0812.0376 (PRL submitted)
\bibitem[\protect\citeauthoryear{Dunkley \etal}{2009}]{dun09}
Dunkley J. \etal\ 2009, ApJS, 180, 306
\bibitem[\protect\citeauthoryear{Enqvist}{2008}]{enq08}
Enqvist K., 2008, Gen. Rel. Grav., 40, 451
\bibitem[\protect\citeauthoryear{Frieman, Turner \& Huterer}{Frieman \etal}{2008}]{fri08}
Frieman J.~A., Turner M.~S., Huterer D., 2008, ARAA, 46, 385
\bibitem[\protect\citeauthoryear{Gupta, Saini \& Laskar}{Gupta \etal}{2008}]{gup08}
Gupta S., Saini T.~D., Laskar T., 2008, MNRAS, 388, 242
\bibitem[\protect\citeauthoryear{Hicken \etal}{2009}]{hic09}
Hicken M. \etal\ 2009, ApJ, 700, 331
\bibitem[\protect\citeauthoryear{Hinshaw \etal}{2009}]{hin09}
Hinshaw G. \etal\ 2009, ApJS, 180, 225
\bibitem[\protect\citeauthoryear{Ishibashi \& Wald}{Ishibashi \& Wald}{2006}]{ish06}
Ishibashi A., Wald R.~M., 2006, CQG, 23, 235
\bibitem[\protect\citeauthoryear{Jha, Riess \& Kirshner}{Jha \etal}{2007}]{jha07}
Jha S., Riess A.~G., Kirshner R.~P., 2007, ApJ, 659, 122
\bibitem[\protect\citeauthoryear{Koivisto \& Mota}{Koivisto \& Mota}{2006}]{koi06}
Koivisto T., Mota D.~F., 2006, Phys. Rev. D, 73, 083502
\bibitem[\protect\citeauthoryear{Koivisto \& Mota}{Koivisto \& Mota}{2008}]{koi08}
Koivisto T., Mota D.~F., 2008, ApJ, 679, 1
\bibitem[\protect\citeauthoryear{Kolatt \& Lahav}{Kolatt \& Lahav}{2001}]{kol01}
Kolatt T.~S., Lahav O., 2001, MNRAS, 323, 859
\bibitem[\protect\citeauthoryear{Kowalski \etal}{2008}]{kow08}
Kowalski M. \etal\ (The Supernova Cosmology Project) 2008, ApJ, 686, 749
\bibitem[\protect\citeauthoryear{Kwan, Francis \& Lewis}{Kwan \etal}{2009}]{kwa09}
Kwan J., Francis M.~J., Lewis G.~F., 2009, arXiv: 0902.4249 (MNRAS accepted)
\bibitem[\protect\citeauthoryear{Leith, Ng \& Wiltshire}{Leith \etal}{2008}]{lei08}
Leith B.~M., Ng S.~C.~C., Wiltshire D.~L., 2008, ApJ, 672, L91
\bibitem[\protect\citeauthoryear{Lynden-Bell \etal}{Lynden-Bell \etal}{1988}]{lyn88}
Lynden-Bell D. \etal\ 1988, ApJ, 326, 19
\bibitem[\protect\citeauthoryear{Maroto}{Maroto}{2006}]{mar06}
Maroto A.~L., 2006, JCAP, 5, 15
\bibitem[\protect\citeauthoryear{Riess \etal}{1998}]{rie98}
Riess A.~G. \etal\ (High-z Supernova Search Team) 1998, AJ, 116, 1009
\bibitem[\protect\citeauthoryear{Riess \etal}{2007}]{rie07}
Riess A.~G. \etal\ (High-z Supernova Search Team) 2007, ApJ, 659, 98
\bibitem[\protect\citeauthoryear{Perlmutter \etal}{1999}]{per99}
Perlmutter S. \etal\ (The Supernova Cosmology Project) 1999, ApJ, 517, 565
\bibitem[\protect\citeauthoryear{Wiltshire}{2007a}]{wil07a}
Wiltshire D.~L., 2007a, New J. Phys., 9, 377
\bibitem[\protect\citeauthoryear{Wiltshire}{2007b}]{wil07b}
Wiltshire D.~L., 2007b, Phys. Rev. Lett., 99, 251101
\end{thebibliography}
\end{document}